\documentclass[prl,twocolumn,superscriptaddress, preprintnumbers]{revtex4}

\usepackage{amsmath}
\usepackage{amsfonts}

\usepackage{graphicx}
\usepackage{times}

\def\gsim{\, \rlap{$>$}{\lower 1.1ex\hbox{$\sim$}}\,}
\def\lsim{\, \rlap{$<$}{\lower 1.1ex\hbox{$\sim$}}\,}

\makeatletter
\renewcommand\section{\@startsection {section}{1}{\z@}%
                                 {-3.5ex \@plus -1ex \@minus -.2ex}
                                   {2.3ex \@plus.2ex}%
                                   {\normalfont\large\bfseries}}
\renewcommand\subsection{\@startsection{subsection}{2}{\z@}%
                                   {-3.25ex\@plus -1ex \@minus -.2ex}%
                                     {1.5ex \@plus .2ex}%
                                     {\normalfont\bfseries}}
\renewcommand\subsubsection{\@startsection{subsubsection}{3}{\z@}%
                                   {-3.25ex\@plus -1ex \@minus -.2ex}%
                                     {1.5ex \@plus .2ex}%
                                     {\normalfont\itshape}}
\makeatother

\newcommand{\be}{\begin{equation}}
\newcommand{\ee}{\end{equation}}
\newcommand{\bea}{\begin{eqnarray}}
\newcommand{\eea}{\end{eqnarray}}
\newcommand{\barr}{\begin{array}}
\newcommand{\earr}{\end{array}}

\def\beq{\begin{equation}}
\def\eeq{\end{equation}}
\def\be{\begin{equation}}
\def\ee{\end{equation}}
\def\bea{\begin{eqnarray}}
\def\eea{\end{eqnarray}}

\begin{document}


\title{Towards a holographic marginal Fermi liquid}

\preprint{SU-ITP-11/24, SLAC-PUB-14442, NSF-KITP-11-066}
\author{Kristan Jensen}
\affiliation{Department of Physics, University of Victoria, Victoria, BC V8W 3P6, Canada}

\author{Shamit Kachru}
\affiliation{Department of Physics, Stanford University and SLAC, Stanford, CA 94305, USA }

\author{Andreas Karch}
\affiliation{Department of Physics, University of Washington,
Seattle, WA  98195, USA}

\author{Joseph Polchinski}
\affiliation{KITP and Department of Physics, UCSB, Santa Barbara, CA 93106, USA}

\author{Eva Silverstein$^2$}

\begin{abstract}
We present an infinite class of 2+1 dimensional field theories which, after coupling to
semi-holographic fermions, exhibit strange metallic behavior in a suitable large $N$ limit.  These theories
describe lattices of hypermultiplet defects interacting with
parity-preserving supersymmetric Chern-Simons theories with $U(N) \times U(N)$ gauge groups at
levels $\pm k$. They have dual gravitational descriptions
in terms of lattices of probe M2 branes in $AdS_4 \times S^7/Z_k$ (for $N \gg  1, N \gg k^5$) or probe D2 branes in $AdS_4 \times CP^3$ (for $N \gg k \gg 1, N \ll k^5$).
We discuss several challenges one faces in maintaining the success of these models at finite $N$, including backreaction of the probes in the gravity solutions and
radiative corrections in the weakly coupled field theory limit.  

\end{abstract}

\maketitle


\paragraph{Introduction}

Local quantum criticality, an invariance under rescaling of energies that leaves the spatial momenta fixed, has been invoked as a potential explanation of interesting phases seen in a variety of condensed matter systems \cite{Si}.    One leading approach for explaining the anomalous transport properties of the strange
metallic phase, the marginal Fermi liquid (MFL) \cite{Varma}, involves a locally critical sector of spin and charge fluctuations, coupled to a Fermi sea.

In general, the theory of non-Fermi liquids is still in its infancy.
One recently developed method of obtaining
controlled models of non-Fermi liquids uses holography.  The study of fermion probes in black brane backgrounds with $AdS_2 \times R^2$
near-horizon geometries \cite{Lee:2008xf,Liu:2009dm,Cubrovic:2009ye,Faulkner:2009wj}, or equivalently
the semi-holographic prescription of \cite{Faulkner:2010tq}, readily gives rise to non-Fermi liquid behavior.
In the latter approach, free fermions are mixed with fermionic operators from a large-$N$ locally critical sector, dual to fermions living in $AdS_2$. 
A distinct holographic mechanism realizing non-Fermi liquid transport arises on probe branes
in Lifshitz backgrounds \cite{Hartnoll:2009ns}.

Much of the work on the holographic approach to non-Fermi liquids has so far been at the level of 4d effective AdS gravity theories, with the scaling dimensions
of operators in the dual field theory appearing as free parameters (masses of bulk fields).  It would be useful to have
microscopic dual pairs where the field theory dynamics giving rise to local criticality is visible in a conventional field
theoretic Lagrangian, and the scaling properties of the non-Fermi liquid can be predicted by the concrete dual field theory
instead of being parameterized as unknowns.\footnote{Other approaches to a microscopic construction of $AdS_2$ duals include Refs.~\cite{Maldacena:1997re,D'Hoker:2009mm}.}
  One goal of our work is to provide an infinite class of such theories where it is natural to obtain precisely the scaling dimensions required for marginal Fermi liquid behavior.

A second goal has been to remedy one of the residual defects in the models of \cite{Liu:2009dm}; there, the precise nature of the  non-Fermi liquid
depends sensitively on the Fermi momentum ${\bf k_F}$ (since the dimensions of the relevant fermionic operators depend on ${\bf k_F}$).
In the models we describe here, the relevant scaling dimension $\Delta$ which (with the right value) gives rise to marginal Fermi
liquid behavior, is independent of ${\bf k_F}$.  This allows an arbitrary shape of the Fermi surface, a useful feature since this is not protected from renormalization
group flow.

A third goal has been to clarify when and how locally critical behavior can occur in a higher-dimensional ($D \geq 2$ dimensional) quantum field theory. Local criticality is a rather exotic property, which needs to be better understood. By definition, it entails quantum mechanical degrees of freedom propagating independently at every point in space, not suppressed by gradient terms. On the other hand, in higher-dimensional quantum field theories, the ultraviolet physics contains itinerant fields which propagate in all directions, with gradient terms in their Lagrangian. Even if one begins with a sector of localized degrees of freedom (like the defects we study), which in itself exhibits local criticality, this sector generically mixes with the itinerant fields through interaction terms. These can, and generally would be expected to, induce gradients. Yet surprisingly, among holographic gravity systems dual to very strongly coupled field theories, one often finds solutions with $AdS_2$ symmetry (using either the AdS-Reissner-Nordstr\"om (RN) black brane, or the world-volumes of appropriate probe branes \cite{Kachru:2009xf} as we shall do here). These solutions are common because they are not terribly hard to obtain, whether by the relatively prosaic matter of stabilizing the extra dimensions of string theory or by stably embedding a probe brane along an $AdS_2$ slice. However even in the large-$N$ approximation of a gauge theory with $N$ colors, strong effects of the itinerant fields are included, so this is a nontrivial result of gauge/gravity duality.

Therefore,
we wish to begin an analysis of whether this emergence of local criticality is only an artifact of the extreme strongly coupled limit where
the gravity description is appropriate, or whether instead a similar mechanism exists also at weaker coupling and finite $N$.  In the second part of this note we discuss the interaction between impurities, which is a finite $N$ effect but becomes important at low energies.  In some cases this spoils the local criticality, but in others this may survive to the IR.

\paragraph{The brane system}

Instead of obtaining $AdS_2$ in the near-horizon limit of an AdS-RN black brane, a setup which
incurs various instabilities, we choose to obtain the $AdS_2$ regions on the worldvolumes of lattice defects, as in
\cite{Kachru:2009xf,Kachru:2010dk}.  A variety of field theoretic toy-models suggest that lattices of defects interacting with itinerant electrons could
be a reasonable starting point for strange metal phenomenology (see e.g. \cite{Varmaagain,Sachdev:2010um,Sachdev}).

Such lattices can be implemented in various ways, differing in their symmetries and in the quantum numbers of the operators in the theory.
The model of \cite{Kachru:2009xf} involves a lattice of defect fermions interacting with the 4d ${\cal N}=4$ supersymmetric
Yang-Mills theory, and is engineered by intersecting D3 and D5-branes (with the D5-branes wrapping $AdS_2 \times S^4$ regions in the near-horizon $AdS_5 \times S^5$ geometry of the D3-branes).   The supersymmetry preserved by that
lattice model is somewhat unconventional (allowing e.g. purely fermionic defect representations); therefore we will
mostly focus on a different lattice system which is 2+1 dimensional and enjoys a more powerful supersymmetry algebra for some values of our discrete parameters.  This, however, entails extraneous bosonic degrees of freedom at the lattice sites, and the examples containing only fermions on the defects can be analyzed similarly.

In the most symmetric case, the brane configuration we study is given, in M-theory, by M2 and M2$^\prime$ branes:
\beq
\label{M2.21} \begin{tabular}{l l l l l l l l l l l l l l}
&$0$ &$1$ &$2$ & \vline &$3$  &4 &5 & 6  &7 &8 &9 & 10\\
  \hline
M2  & x   & x & x &\vline      &      &     &  &  & & & & \\
\hline
M2$^\prime$  &x & :: & :: &\vline  &x &x & &  & &  & & \\
\end{tabular}
\eeq
Here, an ${\rm x}$ denotes a dimension wrapped by the given brane stack, blanks denotes dimensions where
the given branes are localized at a common point, and ::  denotes dimensions in which the given branes are individually localized but form a lattice.  In this
configuration, the two stacks intersect along a lattice
in the 1-2 plane.

Our family of theories will depend on two parameters: $N$ and $k$.  $N$ denotes the number of
M2 branes in the stack above; the M2$^\prime$ branes are equally spaced in a square lattice, and the lattice spacing is
the only scale in the problem (so it doesn't constitute a new parameter).  The second parameter $k$ arises as follows.
We consider a $Z_k$ orbifold which acts as follows on the four complex coordinates transverse to the M2s:
\begin{equation}
\label{orbaction}
g_k: ~~~~ z_i = x_{2i+1} + ix_{2i+2},~~z_i \to {\rm e}^{i 2\pi \over k} z_i,~~i = 1...4 ~.
\end{equation}
The set of M2$^\prime$ branes wrap the locus \cite{Ammon}
\beq
z_1 = z_2 = 0, z_3 = \bar z_4~.
\eeq
and their orbifold images under (\ref{orbaction}). For $k=1$ this embedding is equivalent to the one in (\ref{M2.21}). We treat even and odd $k$ symmetrically, defining the orbifold action to identify points on different, mirror branes (rather than taking the $g_k^{k/2}$ element to identify points on the same brane in the case $k$ even).

The global symmetry of the M2-brane theory is partially broken by the orbifolding and the presence of the M2$^\prime$ probes; from
$SO(8) \times SO(2)$ to $SO(6) \times U(1)\times Z_4$ for $k=1$, and down to $SU(2) \times U(1)^2\times Z_4$ for $k>1$.  The $Z_4$ factor here represents the symmetry of the lattice.
At large $k$ (such that $k^5 \gg N \gg 1$), it follows from the analysis in
\cite{ABJM} that the near-horizon region of the system of M2 and M2$^\prime$ branes is described more accurately using different variables in terms of
type IIA string theory with D2 and D2$^\prime$ branes on a nontrivial geometry with background 2-form gauge flux.


\paragraph{The field theory}

The field theory on the M2 branes in these geometries has been studied in great detail
\cite{ABJM}. 
A general 3d supersymmetric Chern-Simons theory with at least ${\cal N}=2$ supersymmetry has
an action including the terms \cite{GaiottoYin}:
\begin{multline}
\label{basiclag}
S = \int d^3x~{k\over 4\pi} {\rm Tr}(A \wedge dA  + {2\over 3}A^3) + D_{\mu}\bar\phi_i D^{\mu}\phi_i +
i \bar\psi_i \gamma^\mu D_{\mu} \psi_i \\
- {16\pi^2 \over k^2} (\bar\phi_i T^a_{R_i}\phi_i)(\bar\phi_j T^b_{Rj} \phi_j) (\bar\phi_k T^a_{R_k}T^b_{R_k} \phi_k)
\\- {4\pi \over k} (\bar\phi_iT^a_{R_i} \phi_i)(\bar \psi_j T^a_{R_j}\psi_j)- {8\pi \over k} (\bar\psi_i T^a_{R_i}\phi_i)
(\bar\phi_j T^a_{R_j}\psi_j)~.
\end{multline}
Here $T^a_{R}$ are the generators of the gauge group in representation $R$, and the scalars $\phi_i$ and fermions
$\psi_i$ are superpartners in a chiral multiplet.  These terms arise from integrating out the scalars and fermions of the massive vector multiplet and flowing to the deep infrared limit of the theory.

The field theory on our M2 branes is a special case of this theory, with gauge groups $U(N) \times U(N)$ appearing
at levels $\pm k$. The `t Hooft coupling of this theory is $N/k$ and so is large in the holographic limits. The matter fields $\phi_i$ are four bi-fundamental fields $A_{1,2}$ and $B_{1,2}$, in the
$(N, \bar N)$ and $(\bar N, N)$ representations respectively.  In addition to the basic supersymmetric action written above for these
fields, we add an ${\cal N}=3$ superpotential
\beq
W = {2\pi \over k} \epsilon^{ab}\epsilon^{\dot a \dot b} Tr (A_a B_{\dot a}A_b B_{\dot b})~.
\eeq
Here $a,b=1,2$ and the superpotential has been written in a manifestly $SU(2) \times SU(2)$ symmetric manner. The full
symmetry of the field theory is in fact enhanced to an $SO(6) \times U(1)_b$ (with the baryonic $U(1)_b$ acting with charge
$\pm 1$ on the $A$ and $B$ fields), and the theory with these choices enjoys an
enhanced ${\cal N}=6$ supersymmetry \cite{ABJM}.\footnote{In the special cases $k=1,2$,
the supersymmetry is further enhanced to ${\cal N}=8$ and the global symmetry to $SO(8)$.}

The probe M2$^\prime$ branes give rise to localized degrees of freedom; in the type IIA string theory limit of the brane construction these arise from strings stretching between the D2 branes and a lattice of probe D2$^\prime$ branes.
In the simplest case of $k=1$, these are hypermultiplets, with the fermions transforming as spinors in the dimensions transverse to both branes (and the bosons transforming as spinors along 1234).  The infrared Chern-Simons theory is more difficult to analyze directly, since the appropriate type IIB brane construction involves non-perturbative ingredients.  However, by generalizing the methods of \cite{ABJM}\ one can obtain a plausible hypothesis for the spectrum \cite{Ammon}, in which defect hypermultiplets are added to both gauge groups. One reason that this is plausible is that the dual probe branes respect parity, which in the field theory exchanges the gauge group factors. The bosonic quantum mechanical degrees
of freedom $Q_{1,2}$ and $\tilde Q_{1,2}$ at each site transform as follows.  $Q_i$ transforms in the $N$ of the $i$th U(N) gauge group (and is a singlet under the other),
while $\tilde Q_i$ transforms in the conjugate manner; these also transform as spinors under the Lorentz group in the 1234 directions.
Each boson is accompanied by a fermion partner so there are also defect fermions
$\chi_{1,2}$, $\tilde \chi_{1,2}$; these do not transform as spinors in the 1234 directions, but do in the remaining directions.  
Starting from the ABJM theory, the defect probe branes preserve 8 supercharges in the special case of $k=1$, and more generally they preserve 4 supercharges \cite{Ammon}.  We expect a similar spectrum of localized degrees of freedom on the defects for all $k$.

While the overall system preserves at least 4 supercharges in all cases, the superspace structure is unconventional and we have not been able to find a
packaging in the standard superspace arising in 4d ${\cal N}=1$ supersymmetry.  (For instance, from the IIB brane configuration used to obtain the ${\cal N}=6$ theories in \cite{ABJM},
supplemented by our defects as in \cite{Ammon}, it is clear that there are no spatial directions along which one could T-dualize to obtain a higher-dimensional theory with
a conventional superspace; either the probe branes or the ABJM configuration itself breaks the needed higher-dimensional translation symmetries).
However, the couplings of the $A_i, B_j$ fields to the $Q$s and $\tilde Q$s can be inferred by the following logic.  Under translations of the M2 branes along the
$34$ directions, the $Q, \tilde Q$ degrees of freedom should remain massless, while other motions should separate the M2s and M2$^\prime$s and give
$Q, \tilde Q$ a mass.  In a standard way, one can identify motion in the transverse space to the M2 branes with (eigenvalues of) appropriate gauge-invariant composites
of the $A,B$ fields.  First, we identify motion in the 34 directions with $A_1B_1+A_2B_2$.  Then, we expect component couplings localized at the defects depending on the other bilinears in $A_i,B_i$; these are of the form
\begin{multline}
\label{bosonic}
\Delta S = \int dt~\sum_{i} \vert (A_1B_1 - A_2B_2) Q_i\vert^2 + \vert (A_1 B_2 - A_2 B_1)Q_i\vert^2 \\
+ \vert (A_1 B_2 + A_2 B_1) Q_i\vert^2
\end{multline}
with similar couplings to $\tilde Q_i$.  For the fermions, there are related couplings
\be
\label{fermions}
\Delta S = \int dt~\tilde \chi^{\alpha} \Gamma^M_{\alpha\beta} X^M \chi^{\beta}
\ee
with $X^{M}$ corresponding to the real and imaginary parts of $A_1 B_1 - A_2 B_2, A_1 B_2 \pm A_2 B_1$ and $\alpha, \beta$ spinor indices running over the directions transverse to both the M2s and the M2$^\prime$s.

The dimensions of the fields determined from their kinetic terms at
weak coupling are $\Delta(Q) = \Delta (\tilde Q) = -{1\over 2}$,
$\Delta(\chi) = \Delta(\tilde \chi) = 0$, and
$\Delta(A) = \Delta(B) = {1\over 2}$. 
Gauge-invariant composite operators can be formed from these fields.
We will shortly compute the dimensions of low-lying defect operators at strong 't Hooft coupling and large N using the gravity side of the correspondence, and then comment on the field theory description of these operators.



\paragraph{Computation of operator dimensions using holography}

A standard extension of the holographic dictionary relates the dimensions $\Delta$ of scalar operators localized at the lattice sites in our construction,
to the masses of scalar KK modes arising in the M2$^\prime$ brane world-volume action, via the formula
\beq
m^2_{\rm localized} = \Delta (\Delta - 1)~.
\eeq
The fermionic spectrum may be inferred by supersymmetry.

We briefly discuss the calculation in the simplest case, $k=1$.
The fluctuations of the transverse scalars to a given M2$^\prime$ brane (the $x^I=x^5,x^6,..,x^{10}$ directions in space) are all related by an $SO(6)$ symmetry, so
we may focus on a single scalar.
The M2$^\prime$ brane wraps an $AdS_2 \times S^1$ geometry.
The fluctuations can be expanded in Fourier modes on the $S^1$.  If we let $r$ denote
the radial coordinate in $AdS_2$ and focus on static fluctuations, then
\beq
\delta x^I(r,\phi) = \sum_{l} \delta x^{I,l}(r) e^{i l \phi}
\eeq
with $\phi$ the angular coordinate on the wrapped $S^1$.  The resulting Laplace equation for $\delta x^{I,l}(r)$ reveals that
\beq
m_l^2 = -{1\over 4} + {l^2 \over 4}
\eeq
which corresponds to scalar operators of dimension
\beq
\Delta_l = {1\over 2} + {l \over 2}~.
\eeq
The lowest operator in the tower, with $l=0$, gives a sextet of scalar primaries with $\Delta = 1/2$; its Fermi partner is a quartet of $\Delta = 1$ fermionic
defect operators.  We will see in the next subsection that this $\Delta = 1$ multiplet of fermionic operators
plays an important role in obtaining semi-holographic descriptions of marginal
Fermi liquids.

There is also a second tower of operators, arising from fluctuations of the M2$^\prime$ branes along the two transverse spatial directions to their worldvolume in $AdS_4$, i.e. the $x^{1,2}$
directions in (\ref{M2.21}).  The tower arising from these 
fluctuations is distinguished from the tower above by global quantum numbers.  For example, the fluctuations in the $AdS$ directions transform under the $SO(2)$ rotation symmetry of the $x^{1,2}$
plane (which is broken to $Z_4$ by the lattice), and are singlets under the $SO(6)$ global symmetry discussed above, while the fluctuations in the $x^{5,\cdots 10}$ 
directions transform
non-trivially under $SO(6)$ but are $Z_4$ invariant.  While this second tower contains some fermionic operators of $\Delta = 1/2$ which would be dangerous if they coupled
to the semi-holographic fermions, such couplings can be forbidden by the $SO(6) \times Z_4$ symmetry in a ``natural" way (in the sense of the renormalization group).  

The spectrum for higher $k$ may be most easily inferred from the $k=1$ case by the following logic.  We can obtain the higher $k$ brane configurations by $Z_k$
orbifolds of appropriate lattice configurations on $AdS_4 \times S^7$.  The orbifold action is free on the $S^7$ (the fixed point at $z_i=0$ in $C^4$ is removed in the near-horizon
limit), and therefore, all of the low-lying modes in the orbifold theory are $Z_k$ invariant modes in the original $k=1$ theory.
Correlation functions of the dual operators will enjoy large N inheritance from the parent $k=1$ theory, similarly to the theories discussed in \cite{KS}.
 (New degrees of freedom that might be introduced by the orbifolding, analogous to twisted states in string theory, are very massive in the supergravity regime, due to the free orbifold action).   A simple analysis following this logic
implies that the spectrum is the same for all $k>1$; so in particular, $\Delta = 1$ fermionic operators arise in these theories (and any lower $\Delta$ fermionic operators
from the second tower can rendered safe as above, by using global quantum numbers).
A careful discussion of the KK spectra of these theories, and the matching with operators in the dual defect field theories, will appear in \cite{Kristan}.

\paragraph{Coupling to semi-holographic fermions}

The theory we have constructed above is locally critical in the large $N$ limit.  That is, because the probe M2$^\prime$ branes wrap $AdS_2$ slices of the $AdS_4$ geometry,
the excitations of the bulk fields localized on the probe branes can be classified by the quantum numbers of a locally critical quantum theory, and the correlation functions
of the operators dual to localized bulk excitations (computed using the standard AdS/CFT dictionary) obey the constraints following from local criticality.  These are
precisely correlation functions of operators involving defect fields in the dual field theory.

Now, we couple the defect field theory we have constructed to semi-holographic fermions, following \cite{Faulkner:2010tq}.
Namely, if we call the full action of the lattice system above (including both the bulk gauge theory and the defect fields)  ${S}_{LC}$, we consider the theory with
\begin{multline}
\label{actis}
S_{\rm total} = {S}_{LC}(A,B,Q,\tilde Q) +\\
\sum_{J,J'} \int dt~c_J^\dagger (i\delta_{J,J'} \partial_t + \mu \delta_{J,J'} + t_{J,J'}) c_{J'}\\
+ g \sum_{J} \int dt~ (c_J^\dagger {\cal O}^F_J + {\rm Hermitian~conjugate}) ~.
\end{multline}
In (\ref{actis}), we are coupling a normal theory of a weakly coupled Fermi surface (governing the excitations of the $c$ fermion) to the
strongly coupled locally critical sector, through the coupling constant $g$ mixing $c$ with (in any natural theory) the lowest dimension fermionic operator
${\cal O}_F$ that has the right quantum numbers to couple to $c$.

Using large $N$ factorization, it is then easy to show that the $g=0$ Green's function of the $c$ fermion
\beq
G_0({\bf k},\omega) \sim {1\over {\omega - v\vert {\bf k - k_F(k)}\vert}}
\eeq
is modified to
\beq
\label{correctg}
G_{g}({\bf k},\omega) \sim {1 \over {\omega - v \vert{\bf k - k_F(k)}\vert} - g^2 {\cal G({\bf k},\omega)}}~,
\eeq
where
\beq
{\cal G}(\omega) = \int dt ~e^{i\omega t} \langle {\cal O}^{F}_J(t) {\cal O}^{F\dagger}_{J}(0) \rangle~.
\eeq
This two-point function is fixed by the scaling symmetry of the LC theory to be $\mathcal{G}(\omega)=c_{\Delta} \omega^{2\Delta-1}$ where $\Delta$ is the dimension of $\mathcal{O}^F$
(and, importantly, ${\cal G}(\omega) \sim c ~{\omega}~{\rm log} (\omega)$ in the degenerate case $\Delta = 1$).

The correction term in the denominator of $G_g$ will dominate the low-frequency behavior if
$\Delta \leq 1$.  Unitarity allows any $\Delta \geq {1\over 2}$ and this scaling dimension is a free parameter
in the general approaches of \cite{Liu:2009dm,Faulkner:2010tq}.
The marginal Fermi liquid behavior of \cite{Varma} appears in the case that the dimension of
${\cal O}^F$ is precisely 1.  Therefore, the question is, are there natural circumstances in which the theory
$S_{LC}(A,B,Q, \tilde Q)$ has a leading fermionic operator of $\Delta = 1$ which can couple to $c$?

The theories we have constructed above naturally come with defect operators of $\Delta = 1$, as indicated by
our calculation of the KK spectrum on the probe M2$^\prime$ branes.   It is interesting to consider where these come from in field theory language.  The field theory has gauge-invariant
operators of the form
\beq
\partial_t \tilde Q_1 A \chi_2, ~\partial_t  \tilde Q_2 B \chi_1, ~\partial_t Q_1 B \tilde \chi_2, ~\partial_t Q_2 A \tilde \chi_1~.
\eeq
(as well as related quartets of operators of the schematic
form $\tilde \chi_1 \psi_A \chi_2, \cdots$ and $\tilde \chi_1 A \partial_t Q_2, \cdots$).  These have $\Delta=1$ at weak coupling, and are good candidates for the duals of the probe defect operators we computed on the gravity side (arising in the tower of fluctutations of the M2$^\prime$ branes along $x^{5,\cdots,10}$).
Suppose that upon extrapolating to strong coupling (at large N), the weak-coupling dimensions of these operators are  indeed protected, i.e. that the weak-coupling engineering dimensions of the fields correspond to their scaling dimensions under the locally critical scaling governing the defect sector in the probe limit.  Then, assigning appropriate global quantum numbers to $c$, one can choose one of these as the lowest
dimension fermionic operator that $c$ can couple to in the localized sector.   

Returning to the dual gravitational description, we can see that the idea above does work at least in the probe approximation.
By appropriate choice of global quantum numbers (under the $Z_4$ lattice symmetry
and the (subgroup of) $SO(6)$ preserved by the brane configuration), one can guarantee that no lower $\Delta$ operators from the second tower of fluctuations in the
previous subsection infect the leading-order $c$-fermion correlators (\ref{correctg}) after coupling to the large N sector.  
We conclude that we can work directly in the probe limit and obtain a marginal Fermi liquid by identifying $\mathcal{O}^F$ with the lowest fermionic operator in the 
first tower of defect fields computed above.  This has $\Delta=1$, and as emphasized in the introduction, this dimension is independent of momentum.  


\paragraph{Backreaction}

Up until now we have ignored the backreaction of the impurities on the itinerant fields, and therefore on each other.   Thus we have been studying the dynamics of a single impurity interacting strongly with itinerant fields.  The gravity side exhibits the successes it does because the probe branes each wrap an $AdS_2$ region,
and the symmetries of local quantum criticality are manifest, even including the highly nontrivial field theory interactions that are re-summed by the tree-level gravity solution.  

At scales of order the lattice spacing the backreaction is a $1/N$ effect, but at lower energies it must become important.
The scale symmetry of the itinerant fields, which the impurity system inherits, acts on the spatial coordinates.  At energies of order $N^{-1/2}$ times the fundamental scale the number of impurities in a scaling volume is of order $N$, and the effect of the impurities on the itinerant fields and on each other can no longer be neglected.  
Do these effects inevitably generate corrections to the action which destroy the locally critical behavior --- is the behavior seen in the gravity regime a peculiarity of very strongly coupled large $N$
theories, which would not  extrapolate to any more realistic systems --- or can it be robust in some circumstances?
And, if locally critical behavior survives to the far IR, how do the operator dimensions there relate to those we have found at higher energy?

Staying in the limit of strong 't Hooft coupling, gauge/gravity duality transforms this field theory question into the problem of finding the supergravity solution with backreaction.   This can still be a challenging problem, but one can get insight from a simple energetics argument.  We start with the M theory brane configuration~(\ref{M2.21}).  We are looking for an IR geometry $AdS_2 \times R^2 \times X$, which we will for convenience compactify to $AdS_2 \times T^2 \times X$.  We study this with the Ansatz $X = S^7$, averaging the energy density of the impurity $2'$ branes over the compact dimensions.
Let $A$, $T$, and $S$ be the respective radii of the three factors $AdS_2 \times T^2 \times S^7$.
The effective action dimensionally reduced to 1+1 dimensions is of the form
\beq\label{2.2action}
{\cal S}=\int d^2 x \left(- T^2 S^7 + A^2 T^2 S^5 - N'_2 A^2 S - \frac{N_2^2 A^2 T^2}{S^7}\right) \,.
\eeq
We work in units where the M theory scale is one, and ignore order one coefficients.  The respective terms come from the curvatures of $AdS_2$ and $S^7$, the $2'$-brane tensions, and the 7-form flux from the 2-branes.  In other situations it would be natural to Weyl transform to an effective potential, but this is not possible for $AdS_2$; instead we directly extremize with respect to $A$ in addition to $T$ and $S$.  

One finds that there is an extremum (with physically acceptable positive values for the moduli) such that 
\begin{equation}
A\sim S \sim N_2^{1/6}\,,\quad T \sim N_2'^{1/2}/N_2^{1/3} \,. \label{stable}
\end{equation}
The radius S is parametrically the same as for the pure M2 system. The density of defects is $N_2'/T^2 = N_2^{2/3}$.

What is happening is that the lattice defects provide a force acting against the contraction of the two spatial dimensions, hence helping to drive the system towards a fixed point where the bulk modes are locally critical.  
In the probe approximation, the itinerant fields retained their relativistic scaling, and each independent impurity was invariant under a scale transformation leaving its position fixed.  Here there is a common locally critical scaling of the whole geometry.

This result is encouraging, but we should improve the Ansatz.  We have averaged the action of the $2'$ branes over the $S^7$, but in fact they are wrapped on a circle and we should consider moduli corresponding to the contraction of this circle.  Thus we represent $S^7$ as a circle over $CP^3$, with radius $F$ for the fiber circle and $B$ for the base.  The action becomes
\begin{eqnarray}\label{2.2action2}
{\cal S}& =&\int d^2 x \biggl(- T^2 F B^6 + A^2 T^2 F B^4  - A^2 T^2 F^3 B^2 \nonumber\\
&&\qquad\qquad\qquad - N'_2 A^2 F - \frac{N_2^2 A^2 T^2}{F B^6}\biggr) \,.
\end{eqnarray}
One now finds that there is no physical extremum; the contraction of the fiber is not stabilized.

Nevertheless, there are brane systems that realize the solution~(\ref{stable}).  Consider a system with several kinds of impurity brane, with different orientations in the transverse spacetime.  
If the configuration of M$2'$ branes is sufficiently uniform and isotropic, the spherical Ansatz will be a good approximation.\footnote{Such smeared sources have been studied in Ref.~\cite{Nunez:2010sf}.}  Such a configuration will necessarily break supersymmetry (for supersymmetric configurations, at least with ${\cal N} \geq 2$, there will always be an unstable fiber circle).
It is also necessary to stabilize the angular configuration, for example by taking a sufficiently symmetric configuration, and by keeping relatively nonsupersymmetric branes far enough apart to avoid tachyons.  With the scaling~(\ref{stable}) the typical transverse distance between the branes is larger than the M theory scale, so one expects that the latter difficulty may be avoided.  Although with a symmetric distribution there should be a solution of the equations of motion, it may be an unstable saddle point; with the lack of supersymmetry there is no a priori guarantee against disallowed tachyons.
Without having addressed all the possible instabilities, something that might benefit from further model building, we simply take from this construction the lesson already noted that lattice flavors contribute to producing local criticality on the gravity side.




As an aside, the absence of supersymmetric solutions could also be anticipated from another point of view.  We are looking for solutions where the color branes remain localized in the 3-4 directions in which the impurity branes are extended.  In Refs.~\cite{marolf} it is shown that these do not exist for brane intersections of spatial dimension 0 (as here) or 1.  The interpretation was that the scalar fields $Q$ on the intersection are spread out on their  moduli space due to low-dimensional quantum effects, which implies that the brane intersection delocalizes and the $AdS$ IR region disappears.  In nonsupersymmetric systems, masses will generically be generated for these scalars.  
In the appendix we analyze an impurity system that has no such impurity scalars.

Orbifolding by $Z_k$ does not affect the energetics, and so the discussion above can be applied with $N_2 \to N k$, giving in M theory units
\bea
A&\sim& S \sim N^{1/6} k^{1/6}\,, \quad R_{11} \sim N^{1/6} /k^{5/6}\,,\nonumber\\ 
\quad T &\sim& N_2'^{1/2}/N^{1/3}k^{1/3}
\eea
and in string units
\bea
A &\sim& S \sim N^{1/4}k^{1/4} \,,\quad g_s \sim N^{1/4}/k^{5/4} \,, \nonumber\\ T &\sim& N_2'^{1/2} / N^{1/4} k^{1/4}  \,. \label{ABJMscale}
\eea
The same applies if the orbifold action~(\ref{orbaction}) is replaced by one acting only on two complex coordinates $z_{3,4}$, generating the brane configuration
\beq\label{26.2} \begin{tabular}{l l l l l l  l l l l l l l}
&$0$ &$1$ &$2$ & \vline &$3$  &4 &5 & 6  &7 &8 &9\\
  \hline
D2  & x   & x & x &\vline      &      &     &  &  & & & \\
D6  & x   &x  & x &\vline   &x     &x    &x  &x  & & & \\
\hline
D$2'$  &x & :: & :: &\vline  &x &x & &  & &  & \\
\end{tabular}
\eeq
with $N$ color D2-branes and $k$ D6-branes. This is a nice example, having a weakly coupled conformal point for $N_2 \ll N_6$ (as in Refs.~\cite{Appelquist et al}) and an $AdS_4$ dual description for $N_2 \gg N_6$~\cite{Ferrara:1998vf}.  The radius $S$ and coupling $g_s$ are parametrically the same as for the pure D2-D6 system.  In particular one sees that the condition that the radius be large (in string units) is $N_2 \gg N_6$, and that there then is a weakly coupled IIA dual for $N_2 \ll N_6^5$ and an M-theory dual for $N_2 \gg N_6^5$.  The density of defects is $N_2'/T^2 = N_2^{1/2} N_6^{1/2}$.

Even if we find a supergravity solution, there is a general argument that suggests that the local critical scaling cannot persist indefinitely into the IR.  The scaling would imply a density of states 
\begin{equation}
\rho(E) = A \delta(E) + B/E \label{dos}
\end{equation}
 per energy (and exponential in the volume).  The first term is the widely noted zero-temperature entropy.  If only this term is present, the Hamiltonian in the critical sector is zero: there is no dynamics (e.g. a dimension 1 operator would have a correlator $\delta'(t)$ rather than $1/t^2$).  So the $B$ term is necessary, but its integral diverges, so local criticality must always break down at sufficiently low energy.  In the gravity description, the density $B$ comes from bulk states, and so is of order $1/N^2$.  Thus the breakdown takes place at exponentially small scales, which seems more promising than the $N^{-1/2}$ breakdown scale of the probe approximation.

Ref.~\cite{Hartnoll:2009ns} identified a specific breakdown mechanism, whereby the scaling exponents of the spatial directions were shifted (at all scales) from 0 to $O(1/N)$, thus rendering the density of states convergent.  This is a rather special property of the system studied there.  More generally, local criticality might persist until the finite density of states per volume forces it to break down.

\paragraph{Backreaction at weak coupling}

It is encouraging that we have found possible stable systems with the desired IR properties, but the gravity methods are still only controlled in a peculiar limit, from the field theory perspective.
Here we discuss some related issues in direct analysis of the
dual field theory.  We start with the field theory corresponding to the brane system~(\ref{26.2}).  This is 
an ${\cal N}=8$ supersymmetric 3d Yang-Mills theory, with defect hypermultiplets.  In such theories, with a Maxwell action, the conformal symmetry that will emerge in the IR is far from manifest.
A second approach, via the Chern-Simons theories of \cite{ABJM}, has been the one we've followed in the bulk of the paper.
The IR conformal behavior of the bulk theory is much clearer here, as the gauge fields do not appear with a dimensionful coupling,
and the starting (bulk) Lagrangian has no dimensionful parameters.
It is interesting to contrast our expectations for radiative corrections arising from the two approaches.

Starting from the 3d ${\cal N}=8$ Yang-Mills theory with hypermultiplet defects, and following the techniques of \cite{Jay}, it is easy to write a superspace Lagrangian.
The problems with finding a 4d ${\cal N}=1$ superspace do not arise in this perspective; the additional complications of the ABJM brane construction \cite{ABJM}
are not present, and one can straightforwardly T-dualize to find an ${\cal N}=1$ presentation.  In terms of the brane construction with
D2 branes wrapping $x^{1,2}$ and D2$^\prime$ branes
wrapping $x^{3,4}$, it is convenient to perform the T-duality is along the $7,8,9$ directions and to treat those as the spatial directions of the ${\cal N}=1$ field
theory, with $x^{1,2}$ being internal dimensions.
The bulk action is
\begin{multline}
S = {1\over g_3^2} \int dt d^2x ~Tr [ \int d^2\theta {1\over 2}W^{\alpha}W_{\alpha} \\
+ \epsilon^{ijk}\phi_i (\partial_j \phi_k - [\phi_j,\phi_k]/3\sqrt{2}) + h.c.\\
+ 2 \int d^4\theta (\sqrt{2}\bar\partial^i  + \bar\phi^i)e^{-V}(-\sqrt{2}\partial_i + \phi_i)e^{V} + \bar\partial^i e^{-V}\partial_i e^V]\\ + {\rm WZW ~term}~.
\end{multline}
Here, $\partial_1 = \partial_{x^1} + i \partial_{x^2}$, while $\partial_{2,3} \to 0$, and $(\phi^i)^\dagger = \bar\phi^i$.
$W_{\alpha}$ is an $SU(N)$ gauge field strength superfield, while $V$ is the vector superfield.
In 3d ${\cal N}=4$ language, one should think of $\phi_{1,2}$ as the scalars in a hypermultiplet and $\phi_{3}$ as the complex adjoint
scalar in the vector multiplet.
In Wess-Zumino gauge, the WZW term vanishes.   The fields in the above action can be interpreted as follows: D2 gauge field Wilson lines along $x^{1,2}$ and D2 motions along $x^{3,4}$ are packaged in $\phi_{1,2}$;
D2 motions along $x^{5,6}$ are contained in $\phi_3$; and the vector multiplet $V$ has $\theta\bar\theta$ components consisting of $A_0$ and
$x^{7,8,9}$.

The hypermultiplets $H$, which transform in the fundamental of $SU(N)$, have localized actions
\begin{multline}
\sum_{n} \int~dt~\int ~d^4\theta ~(H_n^c e^{V_n}\bar H_n^c + \bar H_n e^{-V_n} H_n) \\
- \int~d^2\theta~ H_n^c \phi_{3,n}H_n - h.c.~.
\end{multline}
The index $n$ runs over the lattice sites, and $n$ subscripts on a bulk field simply indicate that the field is to be evaluated at position of the $n$th site.
This has the intuitively expected features; for instance, motions of the D2 branes along $x^{5,6,7,8,9}$, given the correspondence with
fields above, can be seen to mass up the defect hypermultiplets.

Integrating out the auxiliary $D$-field in the gauge multiplet generates inter-defect interactions.  For simplicity we focus on the Abelian ($N=1$) case;
defect hypermultiplet scalars are denoted by $\eta$.
Then the couplings of the auxiliary field are:
\begin{multline}
S_D = {1\over g_3^2} \int~dt~d^2x~ ({1\over 2}D^2 - 2\sqrt{2}(\phi_1 \bar\partial^1 D + \bar \phi^1 \partial_1 D)\\
+ \dot{ \overline\phi_1} \dot \phi_1) + {1\over 2} \sum_n D_n (\vert \eta^c_n\vert^2 - \vert \eta_n\vert^2)~.
\end{multline}
Integrating out $D$, the action becomes:
\begin{equation}
\label{crossc}
S_D = {1\over g_3^2} \int~dt~d^2x~(-2[\bar\partial^1 {\cal Z}_1 + \partial_1 \bar{\cal Z}^1]^2 + \vert \dot {\cal Z}_1 - \dot\zeta\vert^2)~
\end{equation}
where we've defined
\begin{equation}
\zeta(z_1) = {1\over 8\pi \sqrt{2}} \sum_n {{(\vert \eta_n^c\vert^2 - \vert \eta_n\vert^2)}\over z_1 - z_{1n}}
\end{equation}
and
\begin{equation}
\phi_1 = {\cal Z}_1 - \zeta~.
\end{equation}

The $\vert \dot \zeta\vert^2$ term in (\ref{crossc}) exhibits cross-couplings between the $\eta$ hypermultiplet fields that would naively ruin
local criticality.  One would also get similar terms by integrating out $A_0$ and $\phi_3$.  The generation of inter-defect interactions is not tied to supersymmetry, but these terms sum to a cross-coupling term in the K\"ahler potential for the defect hypermultiplets.
\footnote{There is also a log-divergent same-site kinetic term which we believe can be cancelled by a renormalization of this K\"ahler potential.}
This makes it seem unlikely that the local criticality of the gravity regime can survive to finite $N$ and coupling, where a field
theory analysis should be reliable.
However, it is important to remember that our starting point here has been the 3d ${\cal N}=8$ Yang-Mills theory,
and this UV Lagrangian is valid only far from the IR fixed point which we know governs the physics on the N M2 branes (even at finite $N$).

To get an alternate perspective, we can also try to compute the inter-defect corrections arising from coupling the defect hypermultiplets to the
doubled Chern-Simons theory which captures the fixed-point physics.  In fact, a simple toy-model already illustrates the important difference
between the Chern-Simons defect theories and the Yang-Mills defect theories.  An Abelian Chern-Simons gauge field coupled to defect
fermions $\chi_n$ would be governed by an action
\begin{multline}
S = \int~dt~d^2z~[ A_0 (\partial_z A_{\bar z} - \partial_{\bar z}A_z) - A_z(\partial_0 A_{\bar z} - \partial_{\bar z} A_0)\\
+ A_{\bar z}(\partial_0 A_z - \partial_z A_0) + \sum_n \delta^{(2)}(z-z_n) \chi_n^\dagger A_0 \chi_n]~.
\end{multline}
One can see directly that integrating out $A_0$ will ${\it not}$ generate a dangerous inter-defect coupling here, as it is a non-propagating field.  The $A$ and $B$ fields do propagate, but these couple to the defect fields only quadratically as in Eqs.~(\ref{bosonic}, \ref{fermions}) and so do not generate tree level corrections.

A full field-theoretic analysis of the radiative corrections to the ABJM theory coupled to hypermultiplet defects is beyond the scope
of our work.   It will be interesting to see to what extent the absence of induced inter-defect couplings applies in the full model;
the simple computation above suggests that at least the most obvious dangerous cross-couplings
visible from the Yang-Mills perspective, do ${\it not}$ characterize the physics of the IR fixed point theory coupled to hypermultiplet defects.
Especially in the cases $k=1,2$, where the full model enjoys enhanced supersymmetry, non-renormalization theorems strongly
constrain the possible generation of four-fermion cross-coupling terms (see for instance \cite{AGF}); constraints on higher multi-fermion terms are less obvious.
It would be most interesting to push this analysis further, and construct systems of defect fermions interacting with itinerant fields where local
criticality can be seen robustly directly from field theoretic arguments.

\medskip

\paragraph{Acknowledgments}
We would like to thank S. Kivelson, M. Mulligan, S. Sachdev, and C. Varma for interesting discussions.   We thank S. Hartnoll and J. McGreevy for a correction to the volume dependence of~(\ref{dos}).
K.J., S.K., A.K. and E.S.
acknowledge the hospitality of the Aspen Center for Physics while this work was in progress.  S.K. also acknowledges
the hospitality of the organisers of the 5th Asian Winter School at Jeju Island, and thanks the participants for asking many
interesting questions
about related subjects.  This research was supported in part by the National Science Foundation under grants  PHY-02-44728, PHY05-51164 and PHY07-57035, and by the DOE under contracts DE-AC03-76SF00515 and DE-FG02-96ER40956.  KJ was supported by NSERC, Canada. 
\medskip

\appendix{{\bf Appendix:  The $3.5$ system}

To begin let us consider a variant of the construction of \cite{Kachru:2009xf}, who studied the brane configuration
\beq\begin{tabular}{l l l l l l  l l l l l l l}

&$0$ &$1$  &$2$ &$3$ &\vline  &4 &5  &6 &7 &8 &9\\
  \hline
D3  & x   & x & x & x &\vline      &      &     &  & &   & \\
\hline
D5($\bar 5$)  & x   &:: & :: & :: &\vline   &x     &x    &x  &x &x  & \\
\end{tabular} \label{3.5}
\eeq
As before, an x indicates a direction in which the given branes are extended, and a $::$ indicates a direction in which they are
in a lattice configuration.  The 3-5 intersections are $0+1$ dimensional, representing defects in the dual gauge theory.  For this system, with 8 ND directions, only fermions live on the intersections, which is very natural for the intended applications.  

In the limit that the 5-branes are probes, the D3-branes generate an $AdS_5 \times S^5$ spacetime, with each 5-brane wrapped on an $AdS_2 \times S^4$ subspace.  However, the spatial directions contract in the IR of the $AdS_5$ geometry, so the 5-brane density diverges there and their backreaction cannot be neglected.
At large $N$, the backreaction becomes a large effect at energies which are parametrically small compared to the lattice scale (as noted in \cite{Kachru:2009xf}).
\footnote{Note that the lattice breaks all the conformal symmetries of $AdS_5$: the embedding geometry of each defect is invariant under a different $SO(2,2)$.
The fully backreacted solutions for a single stack of D5 defects will be discussed in \cite{Gonzalo}.}

We are looking for an IR geometry $AdS_2 \times R^3 \times X$, which we will for convenience compactify to $AdS_2 \times T^3 \times X$.  We study this with the Ansatz $X = S^5$, averaging the energy density of the 5-branes over the compact dimensions.  Let $A$, $T$, and $S$ be the respective radii of the three factors $AdS_2 \times T^3 \times S^5$.
The effective action dimensionally reduced to 1+1 dimensions is of the form
\beq\label{3.5action}
{\cal S}=\int d^2 x \left(-\frac{T^3 S^5}{g_s^2}+\frac{A^2 T^3 S^3}{g_s^2} - \frac{{N_5}A^2 S^4}{g_s}-\frac{N_3^2 A^2 T^3}{S^5}\right) \,.
\eeq
We work in units where the string length is one, and ignore order one coefficients.  The respective terms come from the curvatures of $AdS_2$ and $S^5$, the 5-brane tensions, and the RR 5-form flux, and do not distinguish between pure D5-branes and a mix of D5s and $\overline{D5}$s.  In other situations it would be natural to Weyl transform to an effective potential, but this is not possible for $AdS_2$; instead we directly extremize with respect to $A$.   One readily verifies that the action has no stationary points for finite values of the moduli $A,T,S,g_s$.
This analysis precludes an $AdS_2 \times T^3 \times S^5$ solution in the case that the 5-branes are oriented in many directions on the $S^5$, averaging to a symmetric source.


One way of understanding the absence of an $AdS_2$ solution in the infrared in this case is that the ${\cal N}=4$ super Yang-Mills sector has a line of fixed points, parameterized by the string coupling $g_s$.  The additional lattice branes source this mode and altogether there are not enough independent forces to fix $g_s, T, S,$ and $A$.  If we include electric and magnetic flavors, these can fix $g_s$.  Having done this, an $AdS_2$ solution fixing the other moduli does arise.


\begingroup\raggedright\endgroup


\begin{thebibliography}{10}

\baselineskip=14.5pt

\bibitem{Si}
P. Gegenwart, Q. Si and F. Steglich,
``Quantum criticality in heavy fermion metals,''
Nature Physics {\bf 4}, 186 (2008).

\bibitem{Varma}
  C.~M.~Varma, P.~B.~Littlewood, S.~Schmitt-Rink, E. Abrahams, and A.E. Ruckenstein,
  ``Phenomenology of the normal state of Cu-O high-temperature superconductors,''
  Phys.\ Rev.\ Lett.\  {\bf 63}, 1996-1999 (1989).

\bibitem{Lee:2008xf}
  S.~S.~Lee,
  ``A Non-Fermi Liquid from a Charged Black Hole: A Critical Fermi Ball,''
  Phys.\ Rev.\  D {\bf 79}, 086006 (2009)
  [arXiv:0809.3402 [hep-th]].

\bibitem{Liu:2009dm}
  H.~Liu, J.~McGreevy and D.~Vegh,
  ``Non-Fermi liquids from holography,''
  arXiv:0903.2477 [hep-th].

\bibitem{Cubrovic:2009ye}
  M.~Cubrovic, J.~Zaanen and K.~Schalm,
  ``String Theory, Quantum Phase Transitions and the Emergent Fermi-Liquid,''
  Science {\bf 325}, 439 (2009)
  [arXiv:0904.1993 [hep-th]].

\bibitem{Faulkner:2009wj}
  T.~Faulkner, H.~Liu, J.~McGreevy and D.~Vegh,
  ``Emergent quantum criticality, Fermi surfaces, and AdS2,''
  arXiv:0907.2694 [hep-th].

\bibitem{Faulkner:2010tq}
  T.~Faulkner and J.~Polchinski,
  ``Semi-Holographic Fermi Liquids,''
  arXiv:1001.5049 [hep-th].

\bibitem{Hartnoll:2009ns}
  S.~A.~Hartnoll, J.~Polchinski, E.~Silverstein and D.~Tong,
  ``Towards strange metallic holography,''
  JHEP {\bf 1004}, 120 (2010)
  [arXiv:0912.1061 [hep-th]].
  
\bibitem{Maldacena:1997re}
  J.~M.~Maldacena,
  ``The large N limit of superconformal field theories and supergravity,''
  Adv.\ Theor.\ Math.\ Phys.\  {\bf 2}, 231 (1998)
  [Int.\ J.\ Theor.\ Phys.\  {\bf 38}, 1113 (1999)]
  [arXiv:hep-th/9711200].
  
\bibitem{D'Hoker:2009mm}
  E.~D'Hoker, P.~Kraus,
  ``Magnetic Brane Solutions in AdS,''
  JHEP {\bf 0910}, 088 (2009).
  [arXiv:0908.3875 [hep-th]];
  
E.~D'Hoker, P.~Kraus,
  ``Charged Magnetic Brane Solutions in AdS (5) and the fate of the third law of thermodynamics,''
  JHEP {\bf 1003}, 095 (2010).
  [arXiv:0911.4518 [hep-th]];
  
A.~Almuhairi,
  ``AdS$_3$ and AdS$_2$ Magnetic Brane Solutions,''
  [arXiv:1011.1266 [hep-th]].

\bibitem{Kachru:2009xf}
  S.~Kachru, A.~Karch, S.~Yaida,
  ``Holographic Lattices, Dimers, and Glasses,''
  Phys.\ Rev.\  {\bf D81}, 026007 (2010).
  [arXiv:0909.2639 [hep-th]].

\bibitem{Kachru:2010dk}
  S.~Kachru, A.~Karch, S.~Yaida,
  ``Adventures in Holographic Dimer Models,''
  [arXiv:1009.3268 [hep-th]].

\bibitem{Varmaagain}
A.E. Ruckenstein and C.M. Varma,
``A theory of marginal Fermi-liquids,"
Physica C
Vol. 185-189, 134 (1991).



\bibitem{Sachdev:2010um}
  S.~Sachdev,
  ``Holographic metals and the fractionalized Fermi liquid,''
  Phys.\ Rev.\ Lett.\  {\bf 105}, 151602 (2010).
  [arXiv:1006.3794 [hep-th]].

\bibitem{Sachdev}
S.~Sachdev,
  ``Strange metals and the AdS/CFT correspondence,''
  J.\ Stat.\ Mech.\  {\bf 1011}, P11022 (2011)
  [arXiv:1010.0682 [cond-mat.str-el]].

\bibitem{Ammon}
M. Ammon, J. Erdmenger, R. Meyer, A. O'Bannon and T. Wrase,
``Adding flavor to AdS(4)/CFT(3),''
JHEP {\bf 0911} (2009) 125.



\bibitem{ABJM}
O. Aharony, O. Bergman, D. Jafferis and J. Maldacena,
``N=6 superconformal Chern-Simons-matter theories, M2-branes and their gravity duals,''
JHEP {\bf 0810} (2008) 091.

\bibitem{GaiottoYin}
D. Gaiotto and X. Yin, ``Notes on superconformal Chern-Simons-Matter theories,''
JHEP {\bf 0708} (2007) 056.


\bibitem{KS}
S. Kachru and E. Silverstein,
``4d conformal field theories and strings on orbifolds,''
Phys. Rev. Lett. {\bf 80} (1998) 4855.

\bibitem{Kristan}
K. Jensen, to appear.
  
\bibitem{Nunez:2010sf}
  C.~Nunez, A.~Paredes, A.~V.~Ramallo,
  ``Unquenched flavor in the gauge/gravity correspondence,''
  Adv.\ High Energy Phys.\  {\bf 2010}, 196714 (2010).
  [arXiv:1002.1088 [hep-th]].

\bibitem{marolf}
  D.~Marolf, A.~W.~Peet,
  ``Brane baldness versus superselection sectors,''
  Phys.\ Rev.\  {\bf D60}, 105007 (1999).
  [hep-th/9903213].
  
 

\bibitem{Appelquist et al}
  T.~Appelquist and R.~D.~Pisarski,
  ``High-Temperature Yang-Mills Theories And Three-Dimensional Quantum
  Chromodynamics,''
  Phys.\ Rev.\  D {\bf 23}, 2305 (1981);\\
    T.~Appelquist and U.~W.~Heinz,
  ``Vacuum Stability In Three-Dimensional O(N) Theories,''
  Phys.\ Rev.\  D {\bf 25}, 2620 (1982);\\
  T.~Appelquist and U.~W.~Heinz,
  ``Three-Dimensional O(N) Theories At Large Distances,''
  Phys.\ Rev.\  D {\bf 24}, 2169 (1981).

 \bibitem{Ferrara:1998vf}
  S.~Ferrara, A.~Kehagias, H.~Partouche and A.~Zaffaroni,
   ``Membranes and fivebranes with lower supersymmetry and their AdS
  supergravity duals,''
  Phys.\ Lett.\  B {\bf 431}, 42 (1998)
  [arXiv:hep-th/9803109];\\
  J.~Gomis,
  ``Anti de Sitter geometry and strongly coupled gauge theories,''
  Phys.\ Lett.\  B {\bf 435}, 299 (1998)
  [arXiv:hep-th/9803119];\\
  R.~Entin and J.~Gomis,
  ``Spectrum of chiral operators in strongly coupled gauge theories,''
  Phys.\ Rev.\  D {\bf 58}, 105008 (1998)
  [arXiv:hep-th/9804060];\\
  O.~Pelc, R.~Siebelink,
  ``The D2 - D6 system and a fibered AdS geometry,''
  Nucl.\ Phys.\  {\bf B558}, 127-158 (1999).
  [hep-th/9902045].




\bibitem{Jay}
  N.~Marcus, A.~Sagnotti, W.~Siegel,
  ``Ten-dimensional Supersymmetric Yang-mills Theory In Terms Of Four-dimensional Superfields,''
  Nucl.\ Phys.\  {\bf B224}, 159 (1983);

  E.~A.~Mirabelli, M.~E.~Peskin,
  ``Transmission of supersymmetry breaking from a four-dimensional boundary,''
  Phys.\ Rev.\  {\bf D58}, 065002 (1998).
  [hep-th/9712214];
  
  N.~Arkani-Hamed, T.~Gregoire, J.~G.~Wacker,
  ``Higher dimensional supersymmetry in 4-D superspace,''
  JHEP {\bf 0203}, 055 (2002).
  [hep-th/0101233];
  
  N.~R.~Constable, J.~Erdmenger, Z.~Guralnik {\it et al.},
  ``Intersecting branes, defect conformal field theories and tensionless strings,''
  Fortsch.\ Phys.\  {\bf 51}, 732-737 (2003).
  [hep-th/0212265].

\bibitem{AGF}
 L.~Alvarez-Gaume, D.~Z.~Freedman,
  ``Geometrical Structure and Ultraviolet Finiteness in the Supersymmetric Sigma Model,''
  Commun.\ Math.\ Phys.\  {\bf 80}, 443 (1981).


\bibitem{Gonzalo}
S. Harrison, S. Kachru and G. Torroba, work in progress.


\end{thebibliography}
\end{document}